\documentclass{article}
\usepackage{amsmath}
\usepackage{latexsym}

\title{Almost-BPS solutions in multi-center Taub-NUT}
\author{C. ~Rugina$^{a,}$ $^b$ A. Ludu$^c$
\\
\\ \small{a. Department of Theoretical Physics, IFIN-HH, Magurele, Romania,}
\\ \small{b. Department of Physics, University of Bucharest, Bucharest, Romania}
\\ \small{c. Department of Mathematics, Embry-Riddle Aeronautical University, Daytona Beach, FL, USA}
\\}

\begin{document}

\maketitle {\let\thefootnote\relax\footnotetext{{\em Emails}:
 christina.rugina11@alumni.imperial.ac.uk, ludua@erau.edu}}

\begin{abstract}

\noindent
Microstates of multiple collinear black holes embedded in a non-collinear two-center Taub-NUT spacetime are sought in 4 dimensions. A set of coupled ordinary partial differential equations are obtained and solved for almost-BPS states, where some supersymmetry is preserved in the context of $N=2$ supergravity in 4 dimensions. The regularity of solutions is being carefully considered and we ensure that no CTC (closed time-like curves) are present. The larger framework is that of 11-dimensional $N=2$ supergravity and the current theory
is obtained by compactifying down to 4 dimensions. 
\end{abstract}

\section{Introduction}

\noindent
There is a vast literature for BPS solutions for multi-center black holes in 4 dimensions or more \cite{1,2,3,4,5,6,7, 8, 9,10} and also for 5 dimensional black rings \cite{10,11,12,13,14,15}.
While the 4 dimensional black holes solutions are very constrained by uniqueness  theorems, that is not the case for 5-dimensional black rings and we exploit this fact here. We place ourselves with this work in a relatively newer context of finding and classifying non-BPS black holes solutions for two- and three-charge black holes \cite{16,17,18,19,20}. The non-BPS extremal solutions are found as generalizations of the BPS solutions as it was shown for instance in \cite{20}. In the almost-BPS case some supersymmetry survives locally and the equations of the microstates in the supergravity context are found by similarity with the BPS ones. It is well known that introducing supersymmetry makes the complicated Einstein equations tractable and here we, too, are going to separate coupled differential systems of equations and solve them, finding extremal almost-BPS solutions. To be noted though the fact that it seems that extremality is more the key than supersymmetry in solving the differential equations and a lot of the intrinsic features of the solutions are more related to extremality than supersymmetry indeed. To be also noted that the almost-BPS system of differential equations is exactly solvable for generic extremal systems \cite{20} and we shall use this property in this work, too.

\medskip
\noindent
Note that an ensemble of microstates can characterize just like in statistical mechanics the physics of black holes and finding them in a supergravity context can lead to the determination of the black hole's entropy to an acceptable degree of accuracy. The area entropy was derived in the past \cite{21} for classes of 5 and 4 dimensional extremal black holes in string theory by counting the degeneracy of BPS soliton bound states. One can view black holes as BPS soliton solutions which interpolate between maximally symmetric vacua asymptotically and at the horizon. Classically, the entropy of the black hole behaves like a thermodynamic entropy, but the more challenging thing is to give a precise statistical mechanical interpretation of the black hole entropy.
There is a proposal that every black hole microstate should be corrected from a black hole geometry only at the Planck scale and would still have a horizon \cite{22}. Another view is the 'fuzzball' proposal \cite{23} that states that the microstates have strong stringy effects, are smooth solutions, horizonless and that the quantum gravity effects occur at the Schwarzschild radius scale. In this proposal the microstates do not carry entropy and should have the same mass and charges asymptotically as the classical black hole that they describe. Even if the stringy effects are important generic 'fuzzball' microstates can also be described in supergravity and you do get the leading order entropy of the black hole by counting the microstates. One can use though the 'entropy enhancement mechanism' to account for missing black hole entropy in the fuzzball proposal using supertubes (D-brane configurations)\cite{24,25,26}. In magnetically charged backgrounds supertubes see their entropy enhanced by the dipole-dipole interactions with the background \cite{27,28,29,30}.

\medskip
\noindent
There are a number of papers published already that deal with almost-BPS solutions \cite{16,18,19}. The term was actually coined in 2008 \cite{31} and it was used to describe solutions that preserve 1/2 of the supersymmetry locally, but not globally. The most general rotating black hole in 4 dimensions in $N=8$ supergravity was found in \cite{18} and also a non-BPS black ring in Taub-NUT, which descends into 4 dimensions as a two-centered solution: a $\bar{D}$ 6- D2 black hole and the other one is a D4-D2-D0 black hole. In our case we shall study two non-BPS centers (two $\bar{D}$ 6- D2 rotating black holes at the centers of Taub-NUT) - to be noted that in a certain sense this is equivalent to a 5-dimensional non-BPS black ring- and we place a series of collinear BPS D4-D2-D0 black holes, which preserve 1/4 of the supersymmetry in this spacetime. This is now an almost-BPS system. To be noted that there is an interaction $\vec{E} X \vec{B}$ between the BPS black holes that is taken into account by our solution.

\medskip
\noindent
In this context we are going to start with an $N=2$ 11-dimensional supergravity with three families of M2 and M5 M-branes and we are going to postulate a certain ansatz for the metric and the three-form gauge field, as for instance in \cite{15,16,18}. We are then going to compactify down to a 4-dimensional multi-center Taub-NUT (on a 6-torus, comapctification on Calabi-Yau is also possible, but leads to more complicated solutions, although the idea is the same) and we are going to solve the supergravity equations in this context. The solutions are described in terms of harmonic functions, in which instrumental are three anti-self-dual 2-forms describing magnetic fluxes on a hyper-K\"{a}hler 4-dimensional base. The warp factors and the angular momentum are the other unknowns that result naturally from the compactification. We solve the system of equations in which the fact that the magnetic 2-forms are anti-self -dual are instrumental in making the solutions almost-BPS. In the BPS case the magnetic two-forms are self-dual, just like the Riemann tensor. The extra minus sign makes the transition from BPS to almost-BPS solutions, generalizing the first.
The solutions are going to be checked for regularity and the absence of CTCs, the 'bubble' equations fix the positions of the black holes and the 'moduli' of the solutions. They are going to be part of almost-BPS microstates solutions classification and the entropy of the multi-center black holes is going to be determined (calculating the area of the BPS black hole). More recently \cite{19}, generalizations of this type of work have been published in the context of $N=8$ supergravity and a specific case of almost-BPS solutions with one BPS center and many non-BPS centers was treated in detail, in a similar, but more concise and elegant form than that presented here. This same paper recovers the well-known families of almost-BPS solutions derived elsewhere as solutions of differential equations, as nilpotent orbits in simple Lie algebras. They obtain a large class of interacting non-BPS black holes in $N=8$ supergravity (of which $N=2$ is a subcase) with the help of 44 harmonic functions.

\section{Multi-center almost-BPS solutions in multi-center Taub-NUT}

\subsection{The almost-BPS equations}

\bigskip
\noindent
To find the almost-BPS solutions as well as the BPS solutions studied elsewhere[16,18] we start with a 11-dimensional metric carrying M2 and M5 charges and the three-form gauge field sourced by the three families of M2 branes following the ansatz:

\begin{multline}\label{eq:metric11}
ds^2_{11} = -(Z_1 Z_2 Z_3)^{-2/3} (dt + k)^2  + (Z_1 Z_2 Z_3)^{1/3} ds^2_4 + \biggl( \frac{Z_2 Z_3}{Z^2_1}\biggr)^{1/3} (dx^2_1 + dx^2_2) \\ \\ +
\biggl(\frac{Z_1 Z_3}{Z^2_2}\biggr)^{1/3}(d x^2_3 + d x^2_4) +  \biggl(\frac{Z_1 Z_2}{Z^2_3}\biggr)^{1/3}(d x^2_ 5+ d x^2_6),
\end{multline}

\medskip

$$
C^{(3)} = \biggl(a^1 - \frac{dt + k}{Z_1} \biggr)\wedge dx_1 \wedge dx_2 +  \biggl(a^2 - \frac{dt + k}{Z_2}\biggr) \wedge dx_3 \wedge dx_4 
$$
\begin{equation}
+  \biggl(a^3 - \frac{dt + k}{Z_3}\biggr) \wedge dx_5 \wedge dx_6,
\end{equation}

\medskip
\noindent
where $ ds^2_4$ is hyper-K\"{a}hler, four-dimensional metric whose curvature we take to be self-dual. Here $Z_I$ are the warp factors, $\Theta^{(I)} = da^{I}$ are the anti-self-dual dipole field strengths and k is the angular momentum one-form. To be noted that the above metric $(\ref{eq:metric11})$ can be compactified down to 10-dim type IIA supergravity with $C^{(3)}$ as RR field.

\noindent
The almost-BPS solutions are found by analogy with the BPS solutions by changing the sign in the field strengths equation. The almost-BPS equations are:

\begin{equation}
\Theta^{(I)} = -*_{4} \Theta^{(I)},
\end{equation}

\medskip

\begin{equation}
d *_{4} d Z_I = \frac{C_{IJK}}{2}  \Theta^{(I)} \wedge \Theta^{(I)},
\end{equation}

\medskip

\begin{equation}
dk -*_4 dk = Z_I \Theta^{(I)}.
\end{equation}

\medskip
\noindent
The approach of solving this coupled system of equations is to solve first the equations in the angular momentum variable as we shall see in what follows. In this paper we will generalize the results of \cite{16} to non-collinear multi-center case, based on previous "technology" developed in \cite{15, 16, 18}.

\subsection{Solutions with a multi-center Taub-NUT base}

\bigskip
\noindent
We start with a multi-center Taub-NUT base:

\begin{equation}
d^2 s_4 = (V^{m}) ^{-1} (d\psi + A) +V^m ds^2_3
\end{equation}

\medskip
\noindent
with a Gibbons-Hawking potential

\begin{equation}
V^m = h + \frac{q}{r} + \frac{q'}{r'}, \hspace{0.25in} A = q \cos \theta d\phi,  \hspace{0.25in} d^2 s_3 = dr^2 + r^2 d\theta^2 + r^2 \sin^2 \theta d \phi^2.
\end{equation}

\medskip
\noindent
Also $a_i, i= 1, \cdots, N$ are the multiple collinear centers distinct from the non-collinear two-center Taub-NUT origin. At that point $i$, placed at the distance $a_i$ on the $z-$axis, there is a black hole. We introduce the following notations:

\begin{equation}
\Sigma_i = \sqrt{r^2+a^2_i -2 ra_i \cos \theta}
\end{equation}

\medskip
\noindent
and the polar angle of the point i is:

\begin{equation}
\cos \theta_i = \frac{r \cos \theta - a_i}{\Sigma_i}.
\end{equation}

\medskip
\noindent
The magnetic charges (M5) are determined by the harmonic functions $K^{(I)}$:

\begin{equation}
K^{(I)} = \sum_{i=1}^{N} \frac{d^{(I)}_i}{\Sigma_i},
\end{equation}

\medskip
\noindent
where $d^{(I)}_i $ are the magnetic dipoles and I = 1, 2,3. The electric charges (M2) on the other hand are determined by the harmonic functions $L^{(I)}$:

\begin{equation}
L^{(I)} = l_I + \sum_{i=0}^{N} \frac{Q^{(I)}_i}{\Sigma_i}
\end{equation}

\medskip
\noindent
and $Q^{(I)}_i$ are electric charges.

\subsection{Dipole field strengths and warp factors}

\noindent
With the notations introduced in the previous section the two-form field strengths $\Theta^{(I)}$ which are closed and anti-self-dual in the multi-center Taub-NUT spacetime can be written as:

\begin{equation}\label{eq:theta}
\Theta^{(I)} = d [ K^{(I)}(d\psi +A)+ b^{(I)} ],
\end{equation}

\medskip
\noindent
where $b^{(I)}$ obeys the following equation:

\begin{equation}
*_3 db^{(I)} = V^m dK^{(I)} - K^{(I)}dV^m,
\end{equation}

\medskip
\noindent
and so

\begin{equation}\label{eq:b}
b^{(I)} = \sum_i \frac{d^{(I)}_i}{\Sigma_i} \biggl[ h (r \cos \theta - a_i) + q \frac{r - a_i \cos\theta}{a_i} \biggr] d\phi.
\end{equation}

\medskip
\noindent
The warp factors  $Z_i$ which also encode the electric charges (M2) obey the equation:

\begin{equation}
\Box_3 Z_I = V^m \frac{|\epsilon_{IJK}|}{2} \Box_3 (K^{(J)} K^{(K)}) = \biggl( h+\frac{q}{r} +\frac{q'}{r'} \biggr) \sum_{j,k} \frac{|\epsilon_{IJK}|}{2}\Box_3 \biggl( \frac{d^{(J)}_j d^{(K)}_k}{\Sigma_j \Sigma_k} \biggr).
\end{equation}

\medskip
\noindent
Note that:

\begin{equation}
\Box_3 \biggl[ \biggl( \frac{qr}{a_i a_j}+ \frac{q'r'}{a_i a_j} \biggr) \frac{1}{\Sigma_i \Sigma_j} \biggr] = \biggl( \frac{q}{r} +\frac{q'}{r'} \biggr) \Box_3 \biggl( \frac{1}{\Sigma_i \Sigma_j} \biggr).
\end{equation}

\medskip
\noindent
Consequently the complete solution for $Z_I$ is:

\begin{equation}\label{eq:Z}
Z_I = L_I +\frac{|\epsilon_{IJK}|}{2} \sum_{j,k}\biggl( h +\frac{qr}{a_j a_k} +\frac{q' r'}{a_j a_k} \biggr) \frac{d^{(J)}_j d^{(K)}_k}{\Sigma_j \Sigma_k}
\end{equation}.

\subsection{The angular-momentum one-form}

\noindent
We start from the following expression for the angular momentum one-form:

\begin{equation}
k= \mu (d\psi + A) + \omega,
\end{equation}

\medskip
\noindent
where $\mu$ is a scalar function and $\omega$ is a one-form. Then the equation for the angular momentum with the previously introduced notations and functions becomes:

\begin{multline}
d(V^m \mu) + *_3 d\omega = V^m Z_I dK^{(I)} = V^m\sum_{i} l_I d^{(I)}_i d\frac{1}{\Sigma_i}  \\ \\ +
\biggl( h+ \frac{q}{r} +\frac{q'}{r'} \biggr) \sum_{i,j}Q^{(I)}_i d^{(I)}_j \frac{1}{\Sigma_i}d \frac{1}{\Sigma_j} + \frac{|\epsilon_{IJK}|}{2} \sum_{i,j,k}d^{(I)}_i d^{(J)}_j d^{(K)}_k \biggl( h^2 +\frac{hq}{r} \\ \\  +\frac{hqr}{a_j a_k} +
\frac{q^2}{a_j a_k}  + \frac{h q'}{r'} + \frac{h q' r'}{a_j a_k} + \frac{q q' r}{a_j a_k r'} + \frac{q q'r'}{a_j a_k r} + \frac{q'^2}{a_j a_k} \biggr) \frac{1}{\Sigma_j\Sigma_k}d\frac{1}{\Sigma_i}.
\end{multline}

\medskip
\noindent
So the cubic term in $d^{(I)}_i$ is:

\begin{multline}
\sum_{i,j,k}d^{(1)}_i d^{(2)}_j d^{(3)}_k \biggl[ h^2 T^{(1)}_{ijk} + ( q^2 + q'^2 ) T^{(2)}_{ijk} + hq T^{(3)}_{ijk}  \\ \\+
 h' q' T^{'(3)}_{ijk} + q q'\biggl( \frac{r}{r'} T^{(2)}_{ijk}+ \frac{r'}{r} T^{(2)}_{ijk} \biggr) \biggr].
\end{multline}

\medskip
\noindent
One can thus reduce the complete solution for $\mu$ and $\omega$ to
a system of thirteen equations, see Appendix 1, equations (\ref{eq.1} - \ref{eq.3}). Only eight of these thirteen equations with $p$ from $1$ to $8$ (without the primed $4', 5'$ equations) represent the old set of equations for $\mu$ and $\omega$, \cite{16}, with a shift coming from the fact that $V$ is $V^{new}$. 

\noindent

The rest of the $\mu$-s and $\omega$-s  are also obtained in Appendix 1, equations for the index $p=4', 5', 9,\dots, 11$. All these solutions are implemented in the expressions of the potentials, and all the above mentioned contributions are used to build the entire solutions for $\mu$ and $\omega$, equations (\ref{before_final_mu}, \ref{before_final_omega}) in Appendix 1.  With these final values for $\mu$ and $\omega$ the solution can be written completely in terms of angular momentum as:

\begin{equation}
k= \mu (d\psi + A) + \omega.
\end{equation}

\medskip
\noindent
Note that $C^{IJK} = \epsilon_{IJK}$ and that solutions for the warp factor and the anti-self dual dipole fields are given by equations $(\ref{eq:Z})$ and $(\ref{eq:theta})$. k and the solution in general depend on N+6 parameters: $L^{(0)}_{(I)}, K^{(0)}_{(I)}, l_I$, H, q, q', $a_i$, see the section below.

\section{Regularity of solutions}

\subsection{Removing closed time- like curves}

\noindent
We need to impose conditions such that the solution in angular momentum, hence $\omega$  is regular, which indicates the absence of (CTC) - closed-time-like curves and singular Dirac-Misner strings (following prescriptions in \cite{16}). The absence of (CTC) indicates in itself that no time machines can be constructed and that the theory is unitary. We take a look first at the regularity of the $\omega$ solution on the z axis, given that for $\theta = 0 $ or $\pi$ the angle $\phi$ is undefined. In that we require $\omega$ to vanish for these two angles. By checking the various terms contributing, we notice that only $\omega^{(1)}, \omega^{(3)}, \omega^{(5)},  \omega^{(8)}, \omega^{(4')}, \omega^{(5')}, \omega^{(9)}, \omega^{(10)}, \omega^{(11)}$ and $\omega^{(12)}$ are non-vanishing on the z-axis ($\theta=0, \pi $). However, the values for these ten not-null $\omega$-s are different for different $\theta$-s. 

\medskip
\noindent
We present the values of these not-null $\omega$-s at $0$ and at $\pi$ in equations (B.1 - B.8), respectively, in Appendix 2. In the following we denote $s^{(\pm)}_i = \hbox{sign}(r \pm a_i)$. By combining equations (B.1 - B.8) together we obtain the following two conditions in $\theta=0$ and $\theta =\pi$, such that the total solution $\omega$ is null for those two points (which indicates the absence of Dirac-Misner strings):

\begin{multline}
\sum_{i} l_I d^{(I)}_i \frac{s^{(-)}_i}{2} \biggl( h +\frac{q}{a_i} \biggr) + \sum_{i\neq j} Q^{(I)}_i d^{(I)}_j \biggl( h \frac{s^{(-)}_i s^{(-)}_j}{2(a_j - a_i)} + q\frac{s^{(-)}_i s^{(-)}_j}{2 a_j (a_j - a_i)} \biggr) \\ \\+ \sum_{i}Q^{(I)}_i d^{(I)}_i q'  \frac{s^{(-)}_i}{2 a^2_i}  - \sum_{i\neq j} Q^{(I)}_i d^{(I)}_j q' \frac{(a_i + a_j)^2}{2 a^2_i a^2_j} s^{(-)}_i s^{(-)}_j  
\allowdisplaybreaks
\\ \\+  \sum_{i,j,k} d^{(1)}_i d^{(2)}_j d^{(3)}_k
\biggl[ hq \frac{s^{(-)}_i s^{(-)}_j s^{(-)}_k}{2 a_i a_j a_k} + q' h \frac{s_i^{(-)} s_j^{(-)} s_k^{(-)}}{a_i a_j a_k} + q q' \biggl(  \frac{s^{(-)}_j s^{(-)}_k }{ a^2_i a_j a_k} + \frac{s^{(-)}_i s^{(-)}_k}{a_i a^2_j a_k} \\ \\ + \frac{s^{(-)}_i s^{(-)}_j }{a_i a_j a^2_k} + \frac{s^{(-)}_i s^{(-)}_j}{2 a^2_i} + \frac{s^{(-)}_j s^{(-)}_k}{2 a_j a_k} + \frac{s^{(-)}_k s^{(-)}_i}{2 a_k a_i} \biggr) \biggr]  \\ \\ +(\kappa - m_0 -\sum_{i} s^{(-)}_i m_i - \beta) = 0
\end{multline}

\medskip
\noindent
and

\begin{multline}
\sum_{i} l_I d^{(I)}_i \frac{s^{(+)}_i}{2} \biggl( -h +\frac{q}{a_i} \biggr) + \sum_{i\neq j} Q^{(I)}_i d^{(I)}_j \biggl( h \frac{s^{(+)}_i s^{(+)}_j}{2(a_j - a_i)} + q\frac{s^{(+)}_i s^{(+)}_j}{2 a_j (a_j - a_i)} \biggr) \\  \\- \sum_{i}Q^{(I)}_i d^{(I)}_i q'  \frac{s^{(+)}_i}{2 a^2_i}  + \sum_{i\neq j} Q^{(I)}_i d^{(I)}_j q' \frac{(a_i + a_j)^2}{2 a^2_i a^2_j} s^{(-)}_i s^{(-)}_j  \\ \\+  \sum_{i,j,k} d^{(1)}_i d^{(2)}_j d^{(3)}_k
\biggl[ hq \frac{s^{(+)}_i s^{(+)}_j s^{(+)}_k}{2 a_i a_j a_k} - q' h \frac{s_i^{(+)} s_j^{(+)} s_k^{(+)}}{a_i a_j a_k}  \\  \\+ q q' \biggl( \frac{s^{(+)}_j s^{(+)}_k }{a^2_i a_j a_k} + \frac{s^{(+)}_i s^{(+)}_k}{a_i a^2_j a_k} + \frac{s^{(+)}_i s^{(+)}_j }{a_i a_j a^2_k} +  \frac{s^{(+)}_i s^{(-)}_j}{2( a_i a_j)^2} + \frac{s^{(+)}_j s^{(-)}_k}{2( a_j a_k)^2} + \frac{s^{(+)}_k s^{(-)}_i}{2 (a_k a_i)^2} \biggr) \biggr]   \\  \\+ (\kappa + m_0 + \sum s^{(+)}_i m_i - \beta) = 0.
\end{multline}

\medskip
\noindent
These two conditions imply $N+3$ independent constraints derived from the fact that the signs of $s^{(\pm)}_i$ go one way or the other. One can then solve for N+3 independent variables: $k$, $m_0$, $\beta$ and $m_i$ with  $ i = 1, \cdots, N$ (remember that $N$ is the number of black holes centers on the $z-$axis). So the result of that is (taking into account the alternative that all the poles lie at the right of the Taub-NUT center such that: $a_1< a_2< \cdots <a_N)$:

\begin{multline}
\kappa = -q \sum_i \frac{l_I d^{(I)}_i}{ 2 a_i} - h \sum_{i \neq j} \frac{Q^{(I)}_i d^{(I)}_j}{2 (a_j - a_i)} + \\  \\+ \sum_{i,j,k} d^{(1)}_i d^{(2)}_j d^{(3)}_k \biggl[ \frac{h q}{2 a_i a_j a_k} + \frac{q' h}{a_i a_j a_k} + q q' \biggl( \frac{1}{a^2_i a_j a_k} + \frac{1}{a_i a^2_j a_k} + \frac{1}{a_i a_j a^2_k} \biggr) \biggr],
\end{multline}

\begin{multline}
m_0 =  -q \sum_i \frac{l_I d^{(I)}_i}{ 2 a_i} - h \sum_{i} \frac{Q^{(I)}_0 d^{(I)}_j}{2 a_i} + q \sum_{i \neq j, i \neq 0}\frac{Q^{(I)}_i d^{(I)}_j}{2 a_j (a_j - a_i)}  \\  \\+ \sum_{i,j,k} d^{(1)}_i d^{(2)}_j d^{(3)}_k \biggl[ \frac{hq} {2 a_i a_j a_k} + q q' \biggl( \frac{1}{2( a_i a_j)^2} + \frac{1}{2 (a_j a_k)^2} + \frac{1}{2 (a_k a_i)^2} \biggr) \biggr],
\end{multline}

\begin{multline}
m_i = \frac{l_I d^{(I)}_i}{2} \biggl( h+ \frac{q}{a_i} \biggr) + Q^{(I)}_i d^{(I)}_i q' \frac{1}{2 a^2_i}  + \sum_{j} \frac{1}{2 |a_i- a_j|} \biggl[ Q^{(I)}_j d^{(I)}_i \biggl( h+ \frac{q}{a_i} \biggr)  \\ \\ - Q^{(I)}_i d^{(I)}_j \biggl( h+ \frac{q}{a_j} \biggr) - Q^{(I)}_i d^{(I)}_j q' \frac{(a_i+a_j)^2}{2 a^2_i a^2_j} \biggr] + \frac{hq}{2}\biggl( \frac{d^{(1)}_i d^{(2)}_i d^{(3)}_i}{a^3_i}  \\  \\+ \frac{|\epsilon_{IJK}|}{2} \frac{d^{(I)}_i}{a_i} \sum_{j,k} \hbox{sign}(a_j -a_i) \hbox{sign}(a_k - a_i) \frac{d^{(J)}_j d^{(K)}_k}{a_j a_k} \biggr) + hq' \biggl( \frac{d^{(1)}_i d^{(2)}_i d^{(3)}_i}{a^3_i} \biggr)  \\  \\+ q q' \frac{|\epsilon_{IJK}|}{2} \frac{d^{(I)}_i}{a^2_i}\sum_{j,k} \hbox{sign}(a_j -a_i) \hbox{sign}(a_k - a_i) \frac{d^{(J)}_j d^{(K)}_k}{a_j a_k},
\end{multline}

\begin{multline}
\beta = q \sum_i \frac{l_I d^{(I)}_i}{2 a_i} + h \sum_{i \neq j} \frac{Q^{(I)}_i d^{(I)}_j}{2(a_j - a_i)} + \frac{|\epsilon_{IJK}|}{2} \sum_{i,j,k} d^{(I)}_i d^{(J)}_j d^{(K)}_k \biggl( \frac{q q'}{a^2_i a_j a_k} + \frac{hq + hq'}{a_i a_j a_k} \biggr).
\end{multline}

\medskip
\noindent
This way we fix the moduli in the homogeneous solution and we also insure the regularity of the solution at the Taub-NUT center as we are going to see below- the regularity is insured by the above parameters (to be noted that in fact the multi-center Taub-NUT can be thought of as 5-dimensional spacetime as a matter of fact). When there is no black hole at the centers of the Taub-NUT, the metric around r=0, r'=0 is regular. As both $\theta, \phi$ degenerate at this point, regularity requires that $\mu$ and$\omega$ are null at this point. Hence we can verify, given the above relations for the moduli that:

\medskip

\begin{multline}
\mu|_{r,r'=0} = \sum_i \biggl[ l_I d^{(I)}_i \frac{1}{2 a_i} + Q^{(I)}_i d^{(I)}_i \biggl( \frac{1}{2 a^2_i} + \frac{q \cos \theta}{2 h a^3_i} - \frac{q' \sin \theta}{2 h a^3_i} \biggr) \biggr]  \\ \\+
\sum_{i \neq j} Q ^{(I)}_i d^{(I)}_j \biggl( \frac{1}{2 a_i a_j} + \frac{q}{h} \frac{1}{2 a^2_j (a_i - a_j)} \biggr) + \sum_{i,j,k} d^{(1)}_i d^{(2)}_j d^{(3)}_k \biggl[ \frac{q}{a_i a_j a_k}  \\ \\ +
\frac{q' \sin \theta \cos^2 \theta}{a_i a_j a_k} \biggl( \frac{1}{a^2_i} + \frac{1}{a^2_j} + \frac{1}{a^2_k} \biggr) \biggr] + \frac{m_0 + \beta}{h} = 0
\end{multline}

\medskip
\noindent
and similarly for $\omega$:

\medskip

\begin{multline}
\omega |_{r, r'=0} = \biggl[ \sum_i (l_Id^{(I)}_i ) \frac{q \cos \theta-h a_{i}}{2 a_i}  + \sum_{i\neq j} Q^{(I)}_i d^{(I)}_j \biggl( \frac{h}{2 (a_j- a_i)} - \frac{q \cos \theta }{2 a_j(a_i -a_j)}   \\  \\ +
\frac{q'}{2 a_i a_j} \biggr) - \sum_{i, j, k} d^{(1)}_i d^{(2)}_j d^{(3)}_k \frac{h q \cos\theta} {2 a_i a_j a_k} + \kappa - m_0 \cos\theta - \sum_{i \neq 0} m_i - \beta \biggr] d\phi = 0.
\end{multline}

\medskip
\noindent
So we have ensured the regularity of the solution at the centers of Taub-NUT. We can actually verify the regularity of the solution at $r=r'=\theta =0$, as below, taking into account the fact that:

\begin{equation}\label{eq:L0}
L^{(0)}_{(I)} |_{r, r'=0}= l_I + \sum_i\frac{Q^{(I)}_i}{a_i}
\end{equation}

\medskip
\noindent
and

\begin{equation}\label{eq:K0}
K^{(0)}_{(I)} |_{r, r'=0} = \sum_i \frac{d^{(I)}_i}{a_i},
\end{equation}

\medskip
\noindent
which are constants the solution depends on. We will now re-write $\mu$ and $\omega$ at $r=r'=\theta =0$ as functions of the above defined constants $(\ref{eq:L0})$ and $(\ref{eq:K0})$, which will facilitate seeing that they are null at that point. After replacing the constants and some arrangements in $m_0$ and $\beta$, and the other terms in $\mu$ we get:

\begin{multline}
\mu|_{r,r', \theta =0} = d^{(I)}L_{(I)}^{(0)} +\frac{q}{h} L^{(0)}_I K^{(0)}_I + \frac{q}{h} \left(L^{(0)}_{(I)} - l_I \right) K^{(0)}_{(I)} +
\frac{q}{h}\sum_{i \neq j} \frac{L^{(0)}_{(I)} K^{(0)}_{(I)}}{a_i - a_j} \\ \\ + q \sum_{I, J, K} \epsilon_{I J K} K_{(I)} K_{(J)} K_{(K)} - \frac{1}{2}d^{(I)}L_{(I)}^{(0)} -\frac{q}{h} L^{(0)}_I K^{(0)}_I - \frac{q}{h}\sum_{i \neq j} \frac{L^{(0)}_{(I)} K^{(0)}_{(I)}}{a_i - a_j}  \\ \\ - q \sum_{I, J, K} \epsilon_{I J K} K_{(I)} K_{(J)} K_{(K)} - \frac{1}{2}d^{(I)}L_{(I)}^{(0)} \\ \\  + q q'\sum_{I,J,K}\epsilon_{IJK} K_{(I)} K_{(J)} K_{(K)} \sum_{i,j,k} \epsilon _{i, j, k}\left(\frac{1}{a_i}+ \frac{1}{a_j} + \frac{1}{a_k}\right)  \\ \\
+ \frac{q}{h} l^I K_{(I)}^{(0)} - q q'\sum_{I,J,K}\epsilon_{IJK} K_{(I)} K_{(J)} K_{(K)} \sum_{i,j,k} \epsilon _{i, j, k}\left(\frac{1}{a_i}+ \frac{1}{a_j} + \frac{1}{a_k}\right) =0.
\end{multline}

\medskip
\noindent
To see that $\omega$ is null at r=r'=0, we take from the degenerate solution $\theta =0$ and q=q'=0 and we end up with the obviously null quantity:

\begin{multline}
\frac{1}{h} \omega|_{r=0,r'=0, \theta =0} =  -\sum_i l_I d_i \frac{1}{2} + \sum_{i \neq j} Q^{(I)}_i d^{(I)}_j \frac{1}{2 (a_i - a_j)} -  \sum_{i \neq j}\frac{Q^{(I)}_i d^{(I)}_j}{2 (a_j - a_i)} + \\ \\ +
 \sum_j \frac{Q_j^{(I)} d_j^{(I)}}{2 a_i} + \frac{l_i d^{(I)}_i }{2} -\sum_j\frac{ Q_j^{(I)} d_i^{(I)}}{2(a_i-a_j)} + \sum_j\frac{Q_j^{(I)} d^{(I)}_j}{2 a_i} +  \sum_{i\neq j} \frac{Q^{(I)}_i d_j^{(I)}}{2(a_i - a_j)} = 0.
\end{multline}

\subsection{Regularity at the horizons}

\medskip
\noindent
We now turn to study the regularity of the solutions at the poles on the z-axis, which is always true for generic charges and no too large angular momenta. We introduce the convenient quantity:

\begin{equation}
I_4 = Z_1 Z_2 Z_3 V^m - \mu^2 V^{m 2}
\end{equation}

and so the volume element of the horizon around $\Sigma_i = 0$ is:

\begin{equation}
\sqrt{g_{H,i}} = \Sigma_i ( I_4 \Sigma^2_i \sin^2 \theta_i - \omega^2_\phi)^{1/2}.
\end{equation}

\medskip
\noindent
The near-horizon expansion for the point $ r = r' = \Sigma_0 = 0$ for the quantity $I_4$ is:

\begin{equation}
I_4 \approx \frac{Q^{(1)}_0 Q^{(2)}_0 Q^{(3)}_0 q q' - \alpha^2_0 \cos^2 \theta}{r^4 r'}
\end{equation}

\medskip
\noindent
and

\begin{equation}
\omega_\phi \approx \alpha_0 \frac{\sin^2\theta}{r} - \frac{\beta}{r'^2}
\end{equation}

\medskip
\noindent
and so

\begin{equation}
\sqrt{g_{H,0}} \approx (Q^{(1)}_0 Q^{(2)}_0 Q^{(3)}_0 q q' - \alpha^2_0 - \beta^2)^{1/2} \sin \theta.
\end{equation}

\medskip
\noindent
So the spatial measure (area) of the 4-dimensional horizon of the 5-dimensional black hole is:

\begin{equation}
A_{H,0} = (4 \pi q) (4 \pi) (Q^{(1)}_0 Q^{(2)}_0 Q^{(3)}_0 q q' - \alpha^2_0 - \beta^2),
\end{equation}

\medskip
\noindent
where the charge is that of a 5-charge rotating black hole and $\alpha_0, \beta$ encode the 5-dimensional angular momentum. Now expanding carefully around $\Sigma_i = 0$ one gets:

\begin{equation}
I_4 \approx - 2 \alpha_i d^{(1)}_i d^{(2)}_i d^{(3)}_i \biggl( h+ \frac{q + q'}{a_i} \biggr)^2 \frac{\cos\theta_i}{\Sigma^5_i} + \mathcal{O}(\Sigma^{-4}_i),
\end{equation}

\medskip
\noindent

\begin{equation}
\omega_{\phi} \approx \Sigma^{-1}_i.
\end{equation}

\medskip
\noindent
It follows that for regularity (absence of CTCs outside the horizon):

\begin{equation}
\alpha_i = 0  \hspace{0.5in} ( i \ge 1)
\end{equation}

\medskip
\noindent
with this condition the area of the horizon around $\Sigma_i = 0$ is a black ring of area:

\begin{equation}
A_H = 16 \pi^2 q q' J^{1/2}_4
\end{equation}

\medskip
\noindent
and $J_4$ is the $E_{7(7)}$ quartic invariant given by:

\begin{equation}
J_4 = \frac{1}{2}\sum_{I<J} \hat{d}^{(I)}_i \hat{d}^{(J)}_j Q^{(I)}_i Q^{(J)}_i - \frac{1}{4} \sum_I ( \hat{d}^{(I)}_i)^2 (Q^{(I)}_i)^2 - 2 \hat{d}^{(1)}_i \hat{d}^{(2)}_i \hat{d}^{(3)}_i \hat{m}_i,
\end{equation}

\medskip
\noindent
where

\begin{equation}
\hat{d}^{(I)}_i = \biggl( h + \frac{q+q'}{a_i} \biggr) d^{(I)}_i,  \hspace{0.25in} \hat{m}_i = \biggl( h+ \frac{q + q'}{a_i} \biggr)^{-1} m_i.
\end{equation}

\section{Conclusions}

\bigskip
\noindent
We have constructed almost-BPS multi-center solutions that describe two non-BPS rotating black holes and an arbitrary number of collinear-centered BPS black rings in multi-center Taub-NUT (and we are actually covering three non-collinera objects here). In 4 dimensions the arbitrary number of black rings can be thought of as an arbitrary number of collinear $(D4)^3-(D2)^3-D0$ black holes. We checked the regularity of solutions and we fixed the moduli. Recently extremal multi-center non-BPS solutions have been classified using group nilpotent orbits in 4 or 5 dimensional $N=8$ supergravity of which $N=2$ (our case) is a subcase.

\appendix
\section{Appendix 1}
\numberwithin{equation}{section}
\setcounter{equation}{0}

The solutions for the equation (19) can be obtained by solving a system of 13 equations. The first 7 equations have the form:

\begin{equation}\label{eq.1}
d(V^m \mu ^{(1)}_i) + *_3d\omega ^{(1)}_i = V^n d\frac{1}{\Sigma_i},
\end{equation}

\medskip

\begin{equation}\label{eq.2}
d(V^m \mu ^{(p)}_{\mathcal{I}}) + *_3 d\omega ^{(p)}_{\mathcal{I}} =\frac{1}{\xi(p) \Sigma_{i}} d \frac{1}{\Sigma_{i'}}, 
\end{equation}
\noindent 
where indices are assigned as $\mathcal{I}=i$ and $i'=i$ for $p \in \{ 2, 4, 4' \}$, and $\mathcal{I}=\{ i, j \}$, $i'=j$ for $p \in \{ 3, 5, 5' \}$ with restrictions $ij(i-j) \neq 0$. Also $\xi(p)=1$ if $p=2,3$, $\xi(p)=r$ if $p=4,5$, $\xi(p)=r'$ if $p=4', 5'$. The next 6 equations are:
\begin{equation}\label{eq.3}
d(V^m \mu ^{(p)}_{ijk}) + *_3 d \omega ^{(p)}_{ijk} =T^{(p-5)}_{ijk}, \ \  p \in \{ 6, \dots, 11 \},
\end{equation}
\noindent 
where 

\begin{equation}
T^{(1)}_{ijk} = \frac{1}{2} \sum_{(l,m,n)\in \mathcal{P} (i,j,k)}  \frac{1}{\Sigma_l} \frac{1}{\Sigma_m}  d\frac{1}{\Sigma_n},
\end{equation}

\medskip

\begin{equation}
T^{(2)}_{ijk} = \frac{1}{2} \sum_{(l,m,n)\in \mathcal{P} (i,j,k)}
\frac{1}{a_l a_m \Sigma_l} \frac{1}{\Sigma_m}  d\frac{1}{\Sigma_n},
\end{equation}
where $\mathcal{P} (i,j,k)$ represent all permutations. We also have:
\medskip

\begin{equation}
T^{(3)}_{ijk}=\frac{1}{r} T^{(1)}_{ijk}+r T^{(2)}_{ijk},
\end{equation}
and $T^{(4)}$ is identical with $T^{(3)}$ but in $r'$.
In addition we have:

\begin{equation}
T^{(5)}_{ijk} = \frac{r}{r'} T^{(2)}_{ijk}, \ \ \ T^{(6)}_{ijk} = \frac{r'}{r} T^{(2)}_{ijk},
\end{equation}
and
\medskip
\begin{equation}
V^m = h + \frac{q}{r} + \frac{q'}{r'}, \ \ \ V^u =  h+ \frac{q}{r}.
\end{equation}

\noindent
These thirteen equations found in (\ref{eq.1} - \ref{eq.3}) are split between the old set of equations ($p=1, \dots, 4$, $p=5$, and $p=6, 7, 8$) with a shift term coming from the fact that V is $V^{new}$ and the rest. So the solutions to these eight equations are shifted by terms $\mu^{shift}$, respectively $\omega^{shift}$, which are solutions to the following homogeneous equation:

\begin{equation}
d \biggl[ \frac{q'}{r'} \mu (r, r', \theta, \phi)\biggr] + *_3 d\omega (r, r', \theta, \phi) = 0.
\end{equation}

\noindent
The solution to the shift equation is given by:

\begin{equation}\label{eq.A.10}
\mu^{shift}(r, r',\theta, \phi) = \frac{q' r \cos \theta}{r'^2},
\end{equation}

\begin{equation}\label{eq.A.11}
\omega^{shift}(r, r',\theta, \phi) = \frac{r'^2 \sin \theta}{r} d\phi.
\end{equation}

\noindent 
In the following we solve all thirteen equations described in (\ref{eq.1} - \ref{eq.3}). For example, the solution to equation for $p=4'$ is given by:

\begin{equation}\label{eq.A.12}
\omega^{(4')}_i(r, r',\theta, \phi) = \frac{r - r' \cos \theta}{2 a_i \Sigma^2_i} d\phi,
\end{equation}

\medskip

\begin{equation}\label{eq.A.13}
V^m \mu^{(4')}_i (r, r', \theta, \phi) =  \frac{r' \cos\theta- a_i \sin \theta}{2 a_i^2 \Sigma_i^2}.
\end{equation}

\noindent
Proceeding in the same manner we obtain the solutions for $\omega_{\mathcal{I}}^{(p)} (r, r', \theta,\phi)$ and $V^{m}\mu_{\mathcal{I}}^{(p)} (r, r', \theta,\phi)$ for all the other values  of $p=5', 9, 10, 11$, and their corresponding structures of the multi-index $\mathcal{I}$ described above. We also have to consider  the solutions to the homogeneous equation:

\medskip

\begin{equation}\label{eq.A.14}
\mu^{(12)}(r, r',\theta, \phi) = \frac{1}{V^m} \biggl( m_0+ \sum_{i} \frac{m_i}{\Sigma_i} + \sum_{i} \alpha_i \frac{\cos \theta_i}{\Sigma^2_i}+ \frac{\beta}{r'} \biggr) ,
\end{equation}

\medskip

\begin{equation}\label{eq.A.15}
\omega^{(12)} (r, r', \theta, \phi) = \kappa d \phi - \sum_{i} m_i \cos \theta_i d\phi + \sum_{i} \alpha_i \frac{r^2 \sin^2 \theta}{\Sigma^3_i } d\phi - \frac{\beta}{r'^2} d\phi.
\end{equation}

\noindent
Consequently, the solutions to the equations (\ref{eq.1} - \ref{eq.3}) are the old $\mu$ -s and $\omega$ -s (according to reference \cite{16}) plus the $\mu^{shift}$ respectively $\omega^{shift}$.  Putting together these results we can write the final form for each of the  $V^{u}\mu_{\mathcal{I}}^{(p)}$ and $\omega_{\mathcal{I}}^{(p)}$ terms, for all values of $p$ from $1$ to $11$.

\noindent
The  final expressions for $\mu$ and $\omega$ are given in equations (\ref{before_final_mu}, \ref{before_final_omega}) below, where all the terms are calculated similarly to the examples given in equations (\ref{eq.A.10} - \ref{eq.A.15}):

\begin{multline}\label{before_final_mu}
\mu (r, r', \theta, \phi) = \sum_{i} \biggl[ l_I d^{(I)}_i \mu^{(1)}_i  +  Q^{(I)}_i d^{(I)}_i \biggl( h\mu^{(2)}_i + q\mu^{(4)}_i \biggr) \biggr]+ \\ \\
+\sum_{i\neq j} Q^{(I)}_i d^{(I)}_j(h \mu^{(3)}_{ij} +q\mu^{(5)}_{ij}) +
\sum_{i}Q^{(I)}_i d^{(I)}_i q' \mu^{(4')}_i  + \sum_{i\neq j} Q^{(I)}_i d^{(I)}_j q' \mu^{(5')}_{ij} +\\ \\+\sum_{i,j,k} d^{(1)}_i d^{(2)}_j d^{(3)}_k
\biggl[ h^2 \mu^{(6)}_{ijk}  + (q^2 + q'^2)\mu^{(7)}_{ijk}  +hq \mu^{(8)}_{ijk} \biggr] + \\  \\+  \sum_{i,j,k} d^{(1)}_i d^{(2)}_j d^{(3)}_k \biggl[ q' h \mu^{(9)}_{ijk} + q q'\biggl( \mu^{(10)}_{ijk} + \mu^{(11)}_{ijk} \biggr) \biggr] + \mu_{shift} +\mu^{(12)},
\end{multline}

\medskip

\begin{multline}\label{before_final_omega}
\omega (r, r', \theta, \phi)= \sum_{i} \biggl( l_I d^{(I)}_i \omega^{(1)}_i  +  Q^{(I)}_i d^{(I)}_i q \omega^{(4)}_i \biggr)+  \sum_{i\neq j} Q^{(I)}_i d^{(I)}_j(h \omega^{(3)}_{ij} +q\omega^{(5)}_{ij})+ \\ \\ +
 \sum_{i}Q^{(I)}_i d^{(I)}_i q' \omega^{(4')}_i  + \sum_{i\neq j} Q^{(I)}_i d^{(I)}_j q' \omega^{(5')}_{ij} +\\ \\+
 \sum_{i,j,k} d^{(1)}_i d^{(2)}_j d^{(3)}_k \biggl[ h^2\omega^{(6)}_{ijk}  + (q^2 + q'^2) \omega^{(7)}_{ijk}  +hq \omega^{(8)}_{ijk} \biggr] + \\ \\ +
\sum_{i,j,k} d^{(1)}_i d^{(2)}_j d^{(3)}_k \biggl[ q' h \omega^{(9)}_{ijk} + q q'(\omega^{(10)}_{ijk} + \omega^{(11)}_{ijk}) \biggr] + \omega_{shift} +\omega^{(12)}.
\end{multline}

\medskip
\noindent
Putting things together one can obtain the explicit forms of $\mu$  and $\omega $  in terms of the variables $r, r', \theta, \phi$ and parameters $a_{i}, l_{i}, d_{i}$, nevertheless such substitutions generate too long expressions to be presented here.

\medskip

\section{Appendix 2}
\numberwithin{equation}{section}
\setcounter{equation}{0}

The not-null solutions for $\omega$ are presented below. The first option for the $\pm$ sign represent the solutions for $\theta=0$, while the second option for the sign represent solutions for $\theta=\pi$:

\begin{equation}\label{eq.6.1}
\omega^{(1)}_i = \frac{s^{(\mp)}_i}{2} \biggl(\pm h +\frac{q}{a_i} \biggr) d\phi,
\end{equation}

\medskip

\begin{equation}
\omega^{(3)}_{ij} = \frac{s^{(\mp)}_i s^{(\mp)}_j}{2(a_j - a_i)} d\phi, \ \  \omega^{(5)}_{ij} =\frac{\omega^{(3)}_{ij}}{a_{j}},
\end{equation}

\medskip

\begin{equation}
\omega^{(8)}_{ijk} = \frac{s^{(\mp)}_i s^{(\mp)}_j s^{(\mp)}_k}{2 a_i a_j a_k} d\phi,
\end{equation}

\medskip

\begin{equation}
\omega^{(4')}_i = \pm \frac{s^{(\mp)}_i}{2 a^2_i} d\phi,
\end{equation}

\medskip

\begin{equation}
\omega^{(5')}_{ij} = \mp \frac{(a_i + a_j)^2}{2 a^2_i a^2_j} s^{(\mp)}_i s^{(\mp)}_j d\phi,
\end{equation}

\medskip

\begin{equation}
\omega^{(9)}_{ijk} = 2\omega^{(8)}_{ijk} , \ \ \ \omega^{(10)}_{ijk} = \biggl( \frac{s^{(\mp)}_j s^{(\mp)}_k }{a^2_i a_j a_k} + \frac{s^{(\mp)}_i s^{(\mp)}_k }{a_i a^2_j a_k} + \frac{s^{(\mp)}_i s^{(\mp)}_j }{a_i a_j a^2_k} \biggr) d\phi,
\end{equation}

\medskip

\begin{equation}
\omega^{(11)}_{ijk} = \biggl( \frac{s^{(\mp)}_i  s^{(-)}_j}{2 (a_i a_j)^2} + \frac{s^{(\mp)}_j s^{(-)}_k}{2 (a_j a_k)^2} + \frac{s^{(\mp)}_k s^{(-)}_i}{2( a_k a_i)^2} \biggr) d\phi,
\end{equation}

\medskip

\begin{equation}
\omega^{(12)}_{ijk} = \biggl( \kappa \mp m_0 \mp \sum_{i} s^{(-)}_i m_i - \beta \biggr) d\phi.
\end{equation}

\section*{Acknowledgements}

\noindent
This work was initiated under the supervision of Prof. I. Bena of IPhT, CEA, Saclay and C.R. is thankful for the kind hospitality of CEA, Saclay, where part of  this work was completed. C.R. acknowledges a POS-DRU European fellowship. We thank Prof. N. Warner for introducing C.R. to this subject and Dr. C. Ruef and Prof. M. Vi\c{s}inescu for helpful discussions.

\end{document}